\documentclass[preprint,showpacs,amsmath,amssymb]{revtex4}
\usepackage{mathrsfs}

\usepackage{graphicx,color}
\usepackage{subfigure}
\usepackage{dcolumn}
\usepackage{bm}

\begin{document}


\title{Searching for dark matter via mono-$Z$ boson production at the ILC }

\author{Wan Neng$^a$}
\author{Song Mao$^a$}\email{songmao@ahu.edu.cn}
\author{Li Gang$^a$}
\author{Ma Wen-Gan$^b$}
\author{Zhang Ren-You$^b$}
\author{Guo Jian-You$^a$}

\affiliation{$^a$ School of Physics and Material Science, Anhui University, Hefei, Anhui 230039, P.R.China}
\affiliation{$^b$ Department of Modern Physics, University of Science and Technology of China (USTC),
                  Hefei, Anhui 230026, P.R.China}

\date{\today}

\begin{abstract}
High energy colliders provide a new unique way to determine the microscopic properties of the dark matter (DM).
Weakly interacting massive particles (WIMPs) are widely considered as one of the best DM candidates.
It is usually assumed that the WIMP couples to the SM sector through its interactions with quarks and leptons.
In this paper, we investigate the DM pair production associated with a $Z$ boson in an effective field theory
framework at the International Linear Collider (ILC), which can be used to study the interactions
between the DM and leptons. For illustrative purposes, we present the integrated and differential cross sections
for the $e^+ e^- \rightarrow \chi \bar{\chi} Z$ process, where the $Z$ boson is radiated from the initial state
electron or positron. Meanwhile, we analyze the neutrino pair production in association with a $Z$ boson as
the SM background.

\end{abstract}

 \maketitle

\section{Introduction}
\par
Observational evidence has confirmed the existence of some kind of cold non-baryonic dark matter (DM) which is
the dominant component of matter in our universe \cite{Bertone04}. However, astrophysical observations tell us
nothing about the mass of the DM particle or whether it interacts with the standard model (SM) particles. In the
SM, neutrinos are the only long-lived particles that interact purely via the weak force, but their masses are
too small to explain the large mass component in the universe. Thus, to determine the particle nature of DM
is one of the most important tasks in cosmology and particle physics.

\par
Among the many DM candidates, weakly interacting massive particles (WIMPs) are the most compelling ones. Primarily
this is due to the fact that it offers the possibility to understand the relic abundance of the DM as a natural
consequence of the thermal history of the universe \cite{Feng2008}. Theories that lie beyond the SM include
various extensions of the SM, such as supersymmetry \cite{susy1,susy2,susy3,susy4}, universal extra dimensions
\cite{LED1} and little Higgs \cite{LHT1,LHT2}, which all naturally lead to good candidates for WIMPs and the
cosmological requirements for the WIMP abundance in the universe. In these theoretical frameworks, the WIMP
candidates are often both theoretically well motivated and compelling. However, all of these theories still
lack experimental support, and we cannot determine the new physics theory to which the DM belongs. Additionally,
the first observation of the DM may come from direct or indirect detection experiments, which may not provide
information about the general properties of the DM particle without offering a way to distinguish between the
underlying theories. Thus, model-independent studies of DM phenomenology using effective field theory is
particularly important.

\par
Recently, the observational results favour a light DM with a mass around $10~ {\rm GeV}$ in various experiments.
The DAMA experiment has reported a signal of annual modulation at a high significance level \cite{Bernabei:2010mq}.
This signal is consistent with a DM discovery interpretation with a DM particle of mass $\lesssim 10~ {\rm GeV}$
\cite{Petriello:2008jj,Feng:2008dz}. The CoGeNT and XENON10/100 collaborations have also reported a signal
\cite{Aalseth:2010vx,Aprile:2010um,Sorensen:2010hq} which can be explained by a WIMP in this mass range.
There has been much interest in light DM models (where the DM mass is order of a few GeV)
\cite{Kim:2009ke,Fitzpatrick:2010em,Kopp:2009qt,Kuflik:2010ah,Chang:2010yk,Essig:2010ye,An:2010kc,Andreas:2010dz,
Barger:2010yn,Hooper:2010uy}. The high energy colliders are ideal facilities for searching light WIMPs, since
they are most effective when producing highly boosted light WIMPs. In the case of a WIMP, stability on the order
of the lifetime of the universe implies that pair production must highly dominate over single production, and
precludes the WIMP from decaying within the detector volume. WIMPs therefore appear as missing energy, and can
potentially be observed by searching for visible particles recoiling against DM particles \cite{Birkedal:2004xn,
Beltran:2008xg,Cao:2009uw,Beltran:2010ww,Shepherd:2009sa}.

\par
The International Linear Collider (ILC) \cite{123,126} is a proposed
positron-electron collider that is planned to operate at center of
mass energy up to $500~ {\rm GeV}$ with a potential later upgrade to
$1~ {\rm TeV}$. Compared with the Large Hadron collider (LHC), the
$e^+e^-$ linear collider has a particularly clear background
environment. It may have enough energy to produce WIMPs. On the
other hand, positron-electron collider can play a major role in
providing precision data for understanding the DM. At the ILC, the
DM signal has been studied by directly detecting boson transverse
energy signal, such as mono-photon \cite{5740}. Recently, detection
of the DM with a $Z$ boson at the LHC \cite{3352,0231} has been
studied. In this paper, we investigate the DM pair production
associated with a $Z$ boson at the ILC.

\par
The paper is arranged as follows: In Section II we briefly describe
the related effective field theory and present the
calculation strategy. In Section III, we present some numerical
results and discussion. Finally, a short summary is given in Section
IV.

\vskip 5mm
\section{EFFECTIVE FIELD THEORY AND CALCULATION }
\par
The interactions between the SM and DM sectors are presumably
effected by the exchange of some heavy mediators whose nature we do
not need to specify, but only assume that they are much heavier than
the typical scales. The interactions between the DM and SM leptons
are described by an effective Lagrangian as
\begin{eqnarray}
 \mathcal{L} = {\sum}_{\ell} \left\{\frac{1}{\Lambda^2_{D5}}\bar{\ell} \gamma^\mu \ell \bar{\chi} \gamma_\mu\chi
 + \frac{1}{\Lambda^2_{D8}}\bar{\ell} \gamma^\mu \gamma^5 \ell \bar{\chi} \gamma_\mu \gamma^5 \chi
 + \frac{1}{\Lambda^2_{D9}} \bar{\ell} \sigma^{\mu\nu} \ell \bar{\chi} \sigma_{\mu\nu} \chi
 \right\},
\end{eqnarray}
where $\chi$ is the DM particle assumed to be a Dirac fermion,
$\ell$ represents a lepton, and the effective scales $\Lambda_{D5}$,
$\Lambda_{D8}$ and $\Lambda_{D9}$ parameterize the vector (D5),
axial-vector (D8) and tensor (D9) interactions between the two
sectors, respectively. We will typically consider one type of
interaction to dominate at a time, and will thus keep one $\Lambda$
scale finite while the rest are set to infinity and decoupled.

\par
\begin{figure}
\begin{center}
\includegraphics[width=0.6 \textwidth]{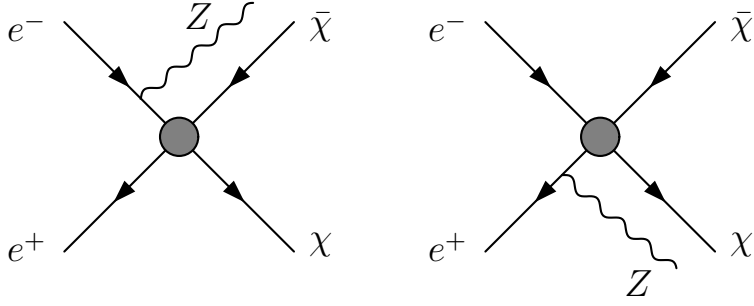}
\caption{ \label{fig1} Tree-level Feynman diagrams for the $e^+e^-
\rightarrow \chi\bar{\chi} Z$ process.}
\end{center}
\end{figure}

There are two Feynman diagrams contributing to the $e^+e^- \rightarrow \chi\bar{\chi} Z$ process at the leading
order (LO), shown in Fig.\ref{fig1}.
The amplitudes for the two diagrams are given by
\begin{eqnarray}
{\cal M}_{i1} = \frac{1}{\Lambda_i^2} \bar{u}(p_4) \Gamma_i v(p_3)
                \bar{v}(p_2) \Gamma^i \frac{i}{\rlap /p_1-\rlap /p_5-m_e}
                \frac{ie\gamma^{\mu}}{4 \sin\theta_W \cos\theta_W}
                (1-4\sin^2\theta_W-\gamma^5) u(p_1)
                \epsilon_{\mu}^*(p_5),~ \nonumber \\
{\cal M}_{i2} = \frac{1}{\Lambda_i^2} \bar{u}(p_4) \Gamma_i v(p_3)
                \bar{v}(p_2) \frac{ie\gamma^{\mu}}{4 \sin\theta_W \cos\theta_W}
                (1-4\sin^2\theta_W-\gamma^5)
                \frac{i}{\rlap /p_5-\rlap /p_2-m_e} \Gamma_i u(p_1)
                \epsilon_{\mu}^*(p_5),~
\end{eqnarray}
where $\Lambda_i = \Lambda_{D5},~ \Lambda_{D8},~ \Lambda_{D9}$ and
$\Gamma_i = \gamma_{\mu},~ \gamma_{\mu}\gamma_5,~ \sigma_{\mu\nu}$
for vector, axial-vector and tensor interactions, respectively.
$p_i~(i = 1,...,5)$ are the four-momenta of the incoming electron, positron
and the outgoing dark matter pair and $Z$ boson, separately. The
differential cross section for the $e^+e^- \rightarrow
\chi\bar{\chi} Z$ process at tree-level is then obtained as
\begin{eqnarray}
\label{cross}
d\sigma_{tree} =\frac{1}{4} \frac{(2 \pi)^4}{2 s} {\sum_{spin}}|{\cal M}_{tree}|^2 d\Phi_3,
\end{eqnarray}
where ${\cal M}_{tree} = {\cal M}_{i1} + {\cal M}_{i2}$ is the
amplitude of all the tree-level diagrams shown in Fig.\ref{fig1}.
The factor $\frac{1}{4}$ is due to taking average over the spins of
the initial particles. $d\Phi_3$ is the three-particle phase space element defined as
\begin{eqnarray}
d\Phi_3=\delta^{(4)} \left( p_1+p_2-\sum_{i=3}^5 p_i \right)
\prod_{j=3}^5 \frac{d^3 \vec{p}_j}{(2 \pi)^3 2 E_j}.
\end{eqnarray}
In our calculations we adopt the 't Hooft-Feynman gauge. The
FeynArts3.4 package \cite{feynarts} is used to generate the Feynman
diagrams and convert them into the corresponding amplitudes. The
amplitude reductions are mainly implemented by employing FormCalc5.4
package \cite{formcalc}.

\vskip 5mm
\section{Numerical results and discussion}
\par
In this section we present the numerical results for the $e^+e^-
\rightarrow \chi\bar{\chi} Z$ process at the ILC. The input
parameters are taken as \cite{pdg}
\begin{eqnarray}
\alpha^{-1}=137.036,~~~ m_Z = 91.1876~{\rm GeV},~~~ m_W = 80.385~{\rm GeV},~~~ m_e=0.511~{\rm MeV}.
\end{eqnarray}
By using the masses of $W$ and $Z$ bosons, we can obtain the value
of weak mixing angle
$\sin^2\theta_W=1-\frac{m_W^2}{m_Z^2}=0.222897$.

\par
For the $e^+e^- \rightarrow \chi\bar{\chi} Z$ process, the final
produced particles $\chi$ and $\tilde{\chi}$ are the missing energy
which will escape the detector without being detected, and the $Z$
boson can be efficiently identified by its leptonic decay. The SM
background mainly comes from the $e^+e^- \rightarrow \nu_{\ell}
\bar{\nu_{\ell}}Z$ $(\ell = e, \mu, \tau)$ processes, where the
neutrino is also the missing energy. Bhabha scattering of leptons
with an additional $Z$ boson, $e^+e^- \rightarrow e^+e^-Z $, is an
important background, which has a large cross section but a very
small selection efficiency, since both final state leptons must be
undetected. As a simple analysis, we don't consider the background
contribution of this part. There are two other important
backgrounds, which are $e^+e^- \rightarrow W^+W^- $ and $W^\pm l^\mp
\nu$ production when the $W$ boson(s) decays leptonically. By
adopting appropriate event selection, these backgrounds can also be
safely ignored \cite{Cheung:1999zw}.

\par
\begin{figure}
\begin{center}
\includegraphics[width=0.6\textwidth]{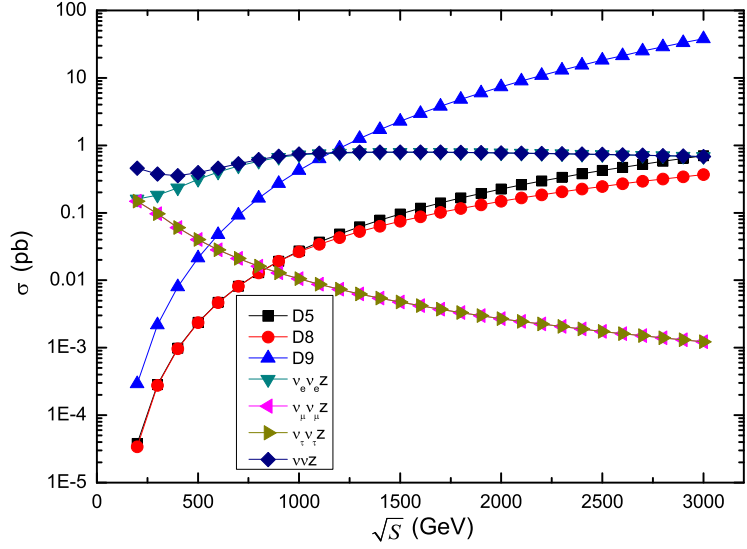}
\caption{ \label{fig2} Total cross sections for the signal process
$e^+e^- \rightarrow \chi\bar{\chi} Z$ induced by the vector (D5),
axial-vector (D8), tensor (D9) operators and the background
processes $e^+e^- \rightarrow \nu_{\ell} \bar{\nu_{\ell}}Z$ $(\ell =
e, \mu, \tau)$ as functions of the colliding energy $\sqrt{s}$ by
taking $m_\chi = 10~ {\rm GeV}$, $\Lambda=1~ {\rm TeV}$ and
$10^{\circ} < \theta_z < 170^{\circ}$. }
\end{center}
\end{figure}

In Fig.\ref{fig2}, we present the cross sections as functions of the
colliding energy $\sqrt{s}$ for the signal induced by the vector,
axial-vector, tensor operators and the background including three
generations of neutrinos by taking $m_\chi = 10~ {\rm GeV}$ and
$\Lambda=1~ {\rm TeV}$ with $10^{\circ} < \theta_z < 170^{\circ}$
(There $\theta_z$ is the angle between the $Z$ boson and
the incoming electron beam.), separately. From this figure
we can see that, with the increment of the colliding energy
$\sqrt{s}$, the cross sections for the signal process $e^+e^-
\rightarrow \chi\bar{\chi} Z$ induced by the vector, axial-vector
and tensor operators increase rapidly, the cross section for the
background from $\nu_e \bar{\nu}_e Z$ production vary slowly, while
those backgrounds from $\nu_{\mu} \bar{\nu}_{\mu} Z$ and $\nu_{\tau}
\bar{\nu}_{\tau} Z$ production processes decrease slightly. With the
increment of $\sqrt{s}$, the ratio of background to signal declines
gradually and the signal becomes significant relative to the
background. Consequently, we can obtain larger cross section for the
signal and improve the probability for searching DM by raising the
colliding energy $\sqrt{s}$.

\par
\begin{figure}
\begin{center}
\includegraphics[width=0.8\textwidth]{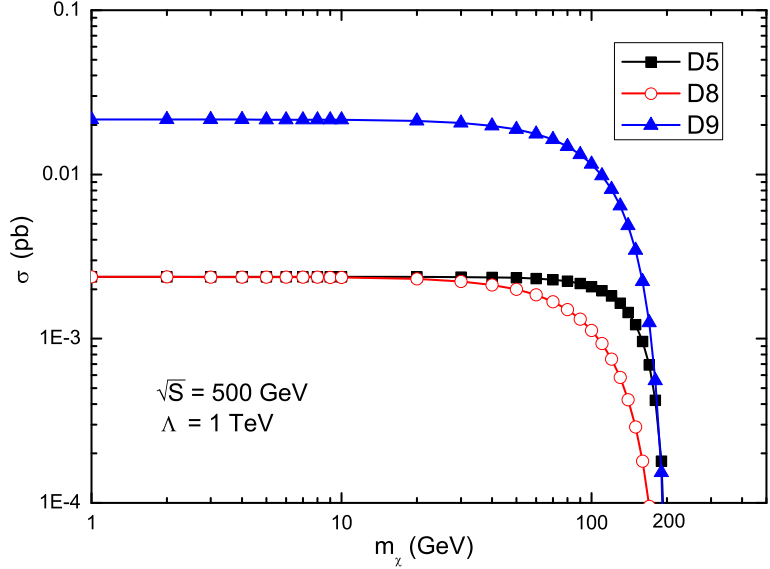}
\caption{ \label{fig3} Dependence of cross sections for the signal
process $e^+e^- \rightarrow \chi\bar{\chi} Z$ induced by the vector
(D5), axial-vector (D8) and tensor (D9) operators on the DM mass
with $\sqrt{s} = 500~{\rm GeV}$, $\Lambda=1~ {\rm TeV}$ and
$10^{\circ} < \theta_z < 170^{\circ}$.}
\end{center}
\end{figure}

In Fig.\ref{fig3}, we present the DM mass dependence of the cross
sections for the $e^+e^- \rightarrow \chi\bar{\chi} Z$ process
induced by the vector, axial-vector and tensor operators by taking
$\sqrt{s}=500~ {\rm GeV}$, $\Lambda=1~ {\rm TeV}$ and $10^{\circ} <
\theta_z < 170^{\circ}$, separately. As shown in this figure, the
cross section is insensitive to the DM mass $m_{\chi}$ in the range
of $m_{\chi} < 100~ {\rm GeV}$, and decreases rapidly with the
increment of $m_\chi$ when $m_{\chi} > 100~ {\rm GeV}$, due to the
rapidly reduced phase space. We also conclude that the contributions
from the spin-independent operator (D5) and spin-dependent operator
(D8) can be distinguished until $m_{\chi} > 100~ {\rm GeV}$.

\begin{figure}
\begin{center}
\includegraphics[width=0.45\textwidth]{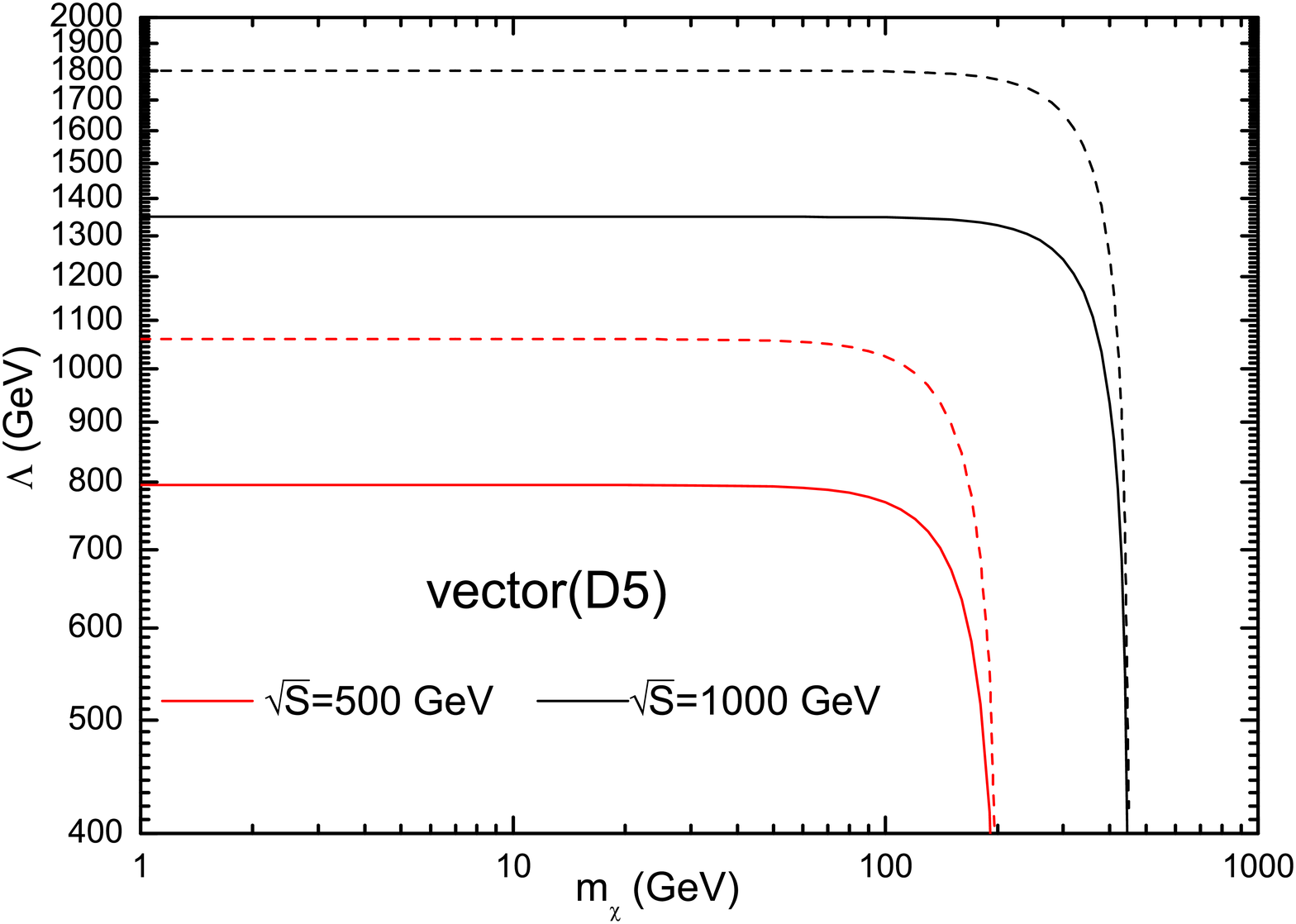}
\includegraphics[width=0.45\textwidth]{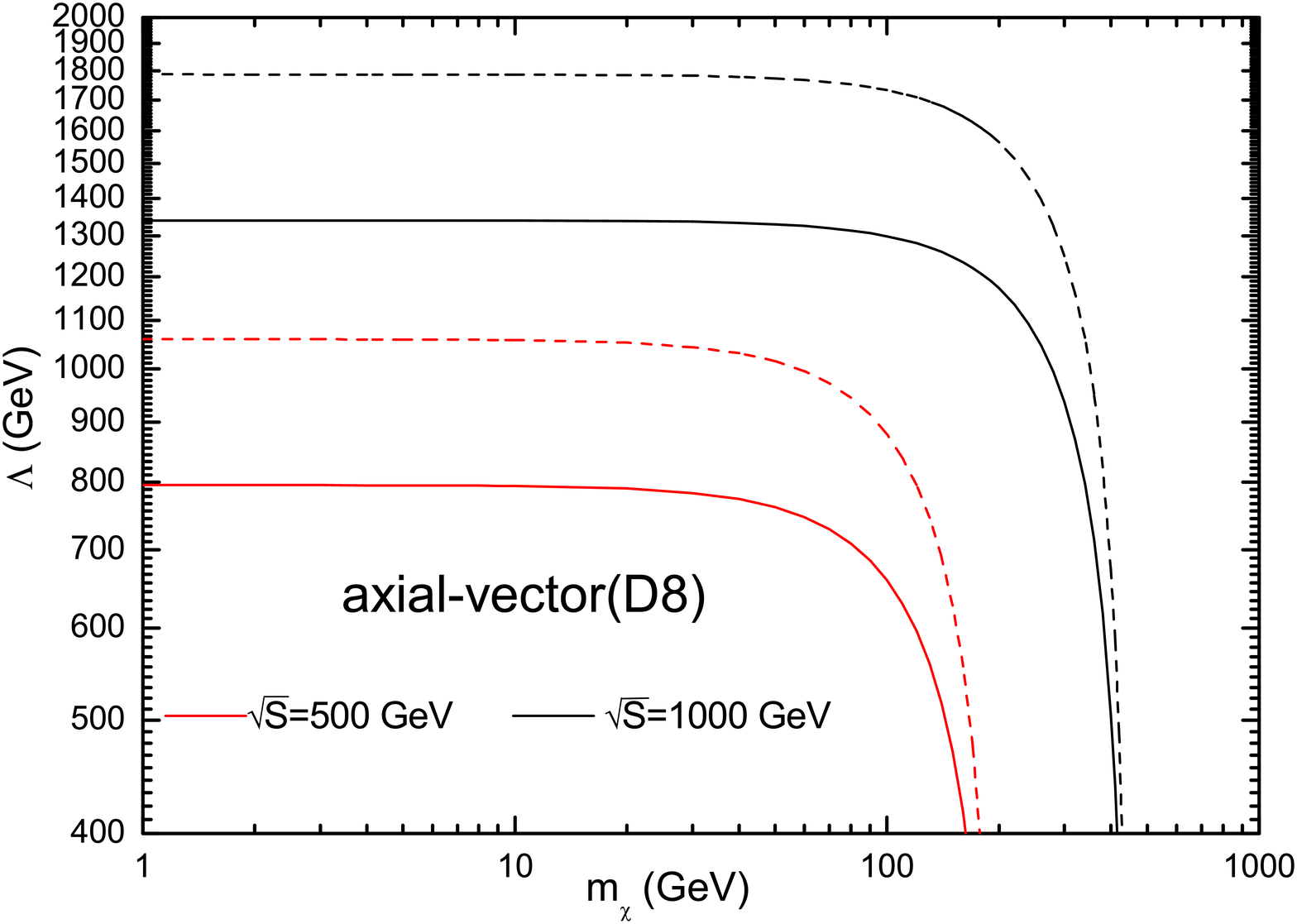}
\includegraphics[width=0.45\textwidth]{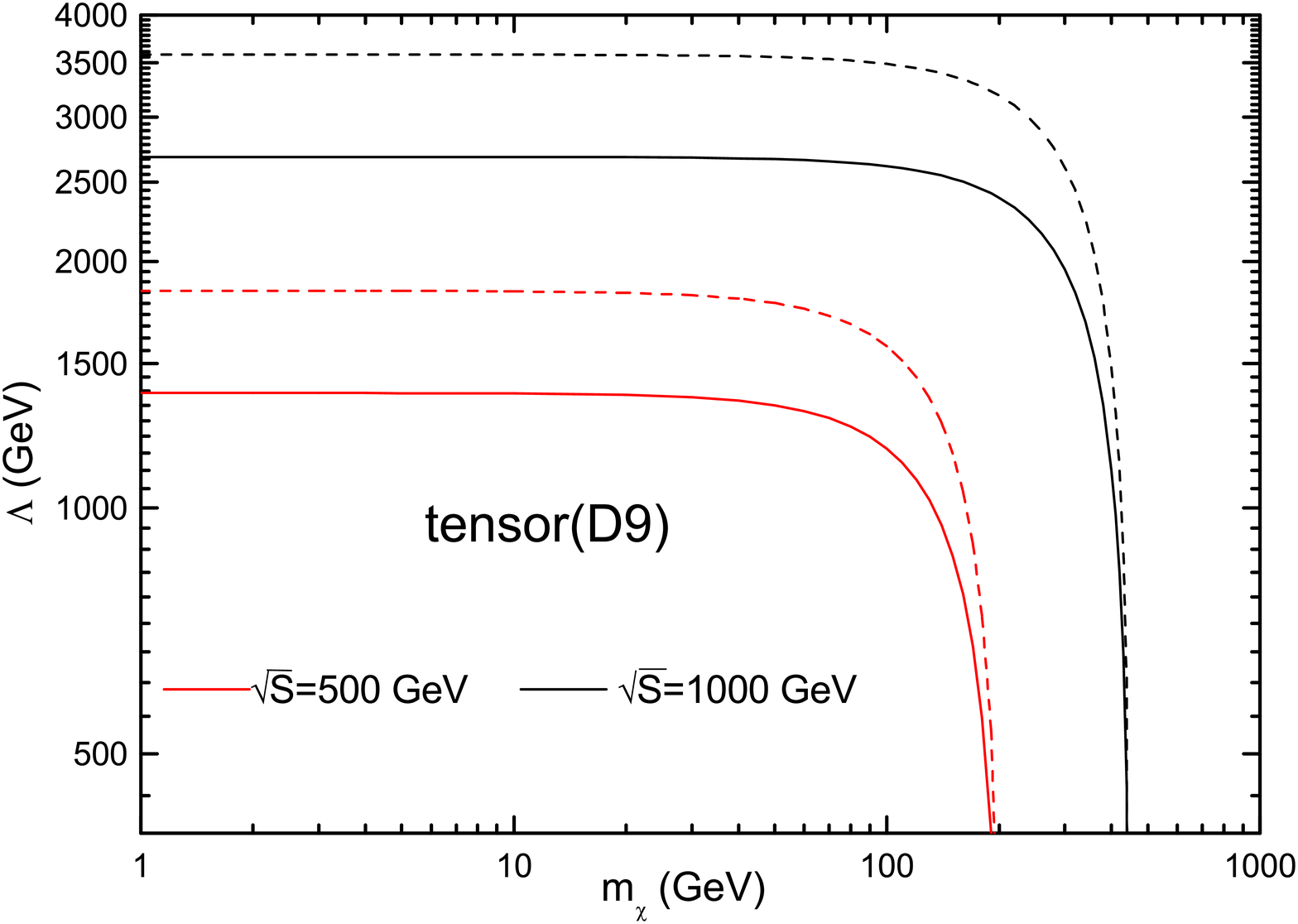}
\caption{ \label{fig4} $3\sigma$ detection region on the
$m_\chi-\Lambda$ plane for the $\chi\tilde{\chi}Z$ production
induced by the vector (D5), axial-vector (D8) and tensor (D9)
operators  at the $\sqrt{s} = 500$ and $1000~{\rm GeV}$ ILC with
integrated luminosities of $100 fb^{-1}$ (solid lines) and $1000
fb^{-1}$(dashed lines), respectively.}
\end{center}
\end{figure}

\begin{figure}
\begin{center}
\includegraphics[width=0.45\columnwidth]{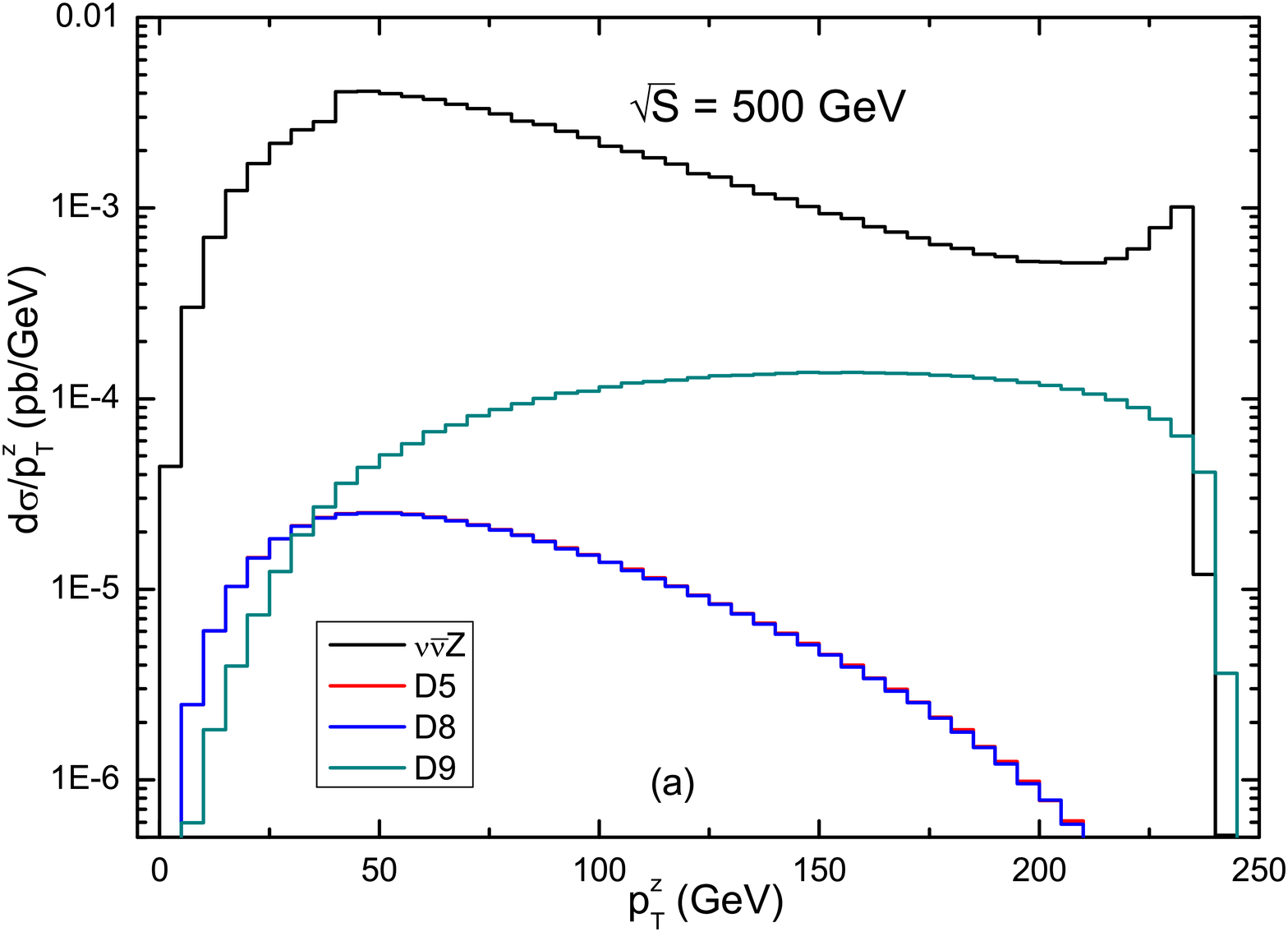}
\includegraphics[width=0.45\columnwidth]{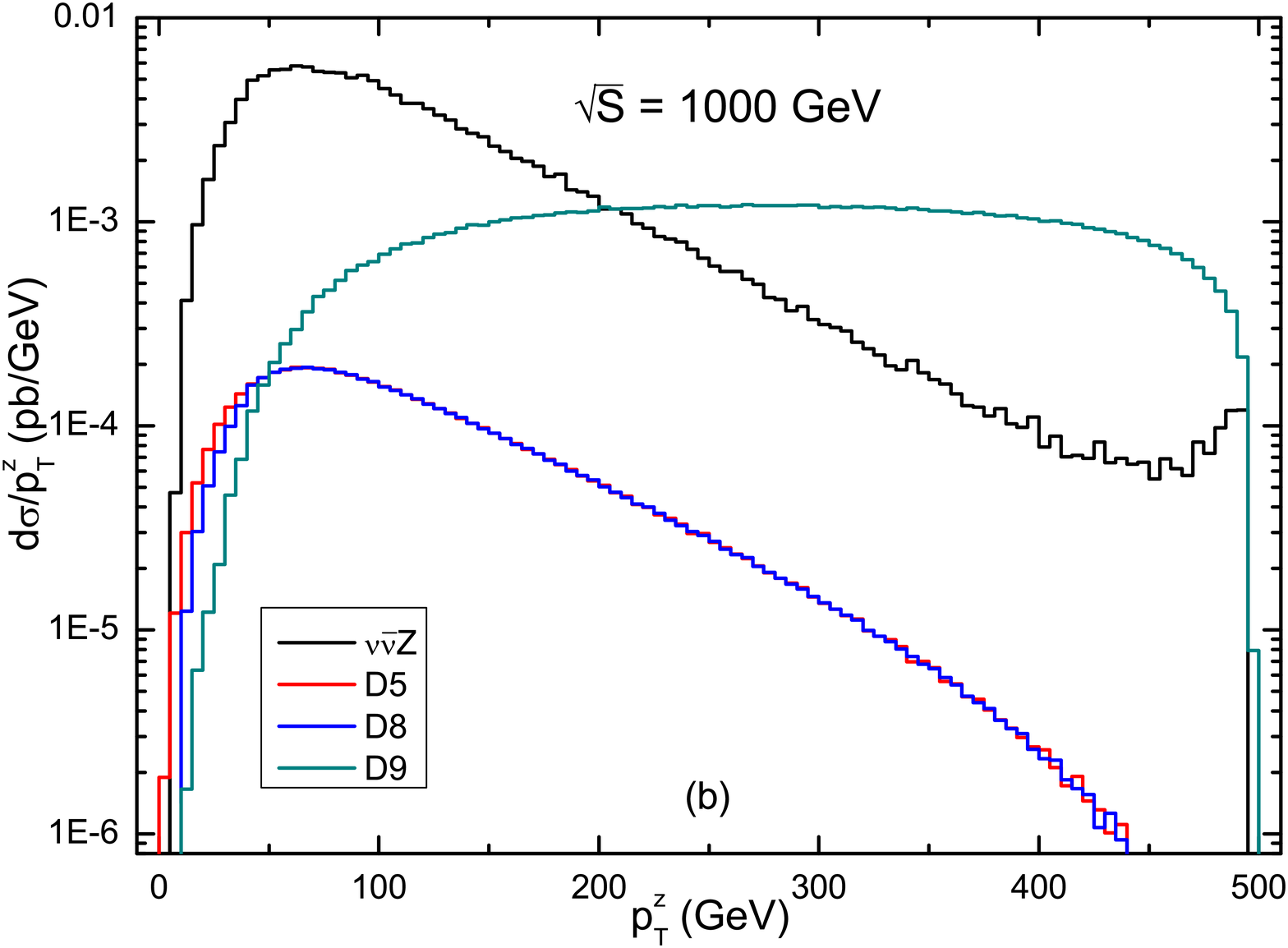}
\includegraphics[width=0.45\columnwidth]{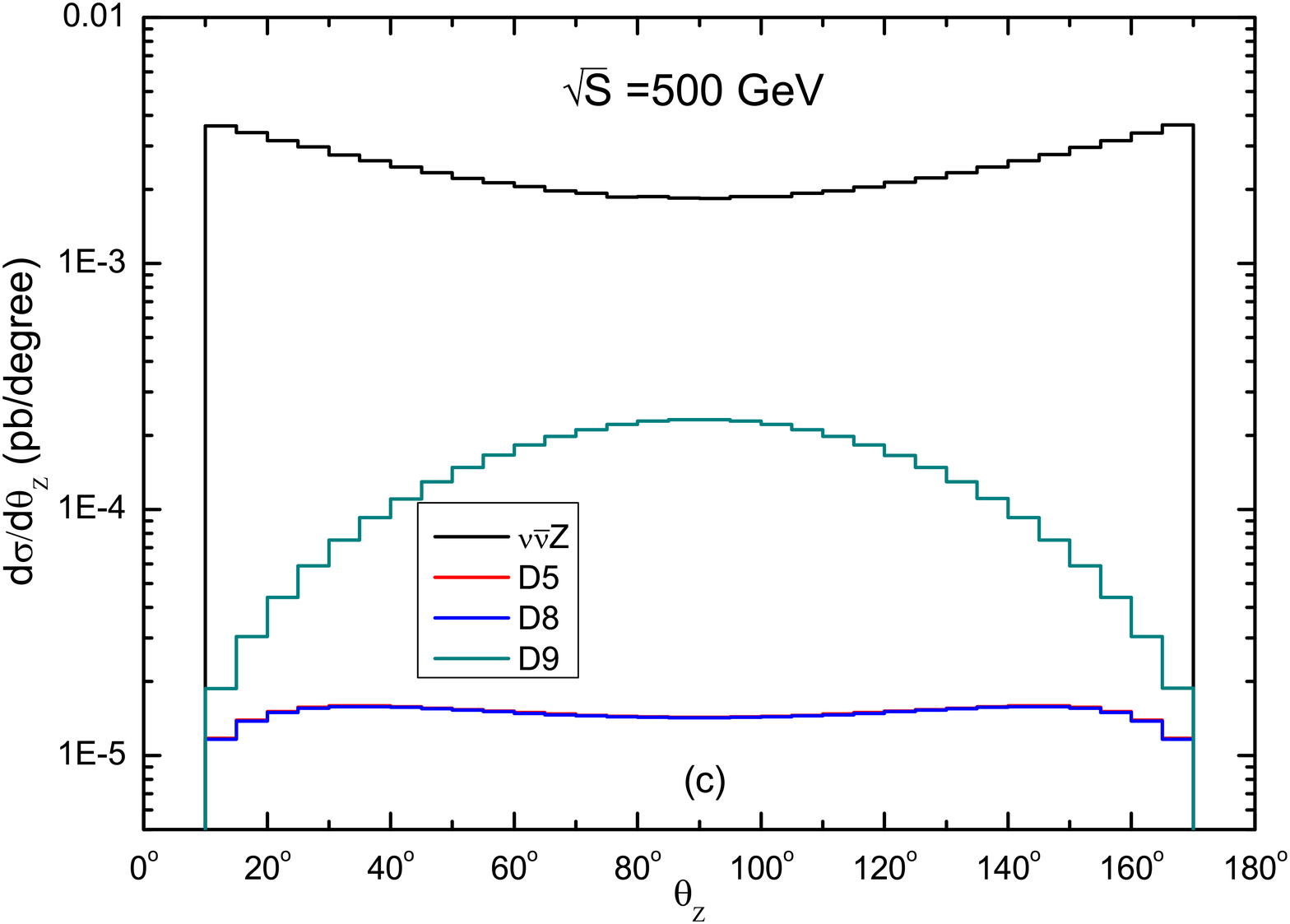}
\includegraphics[width=0.45\columnwidth]{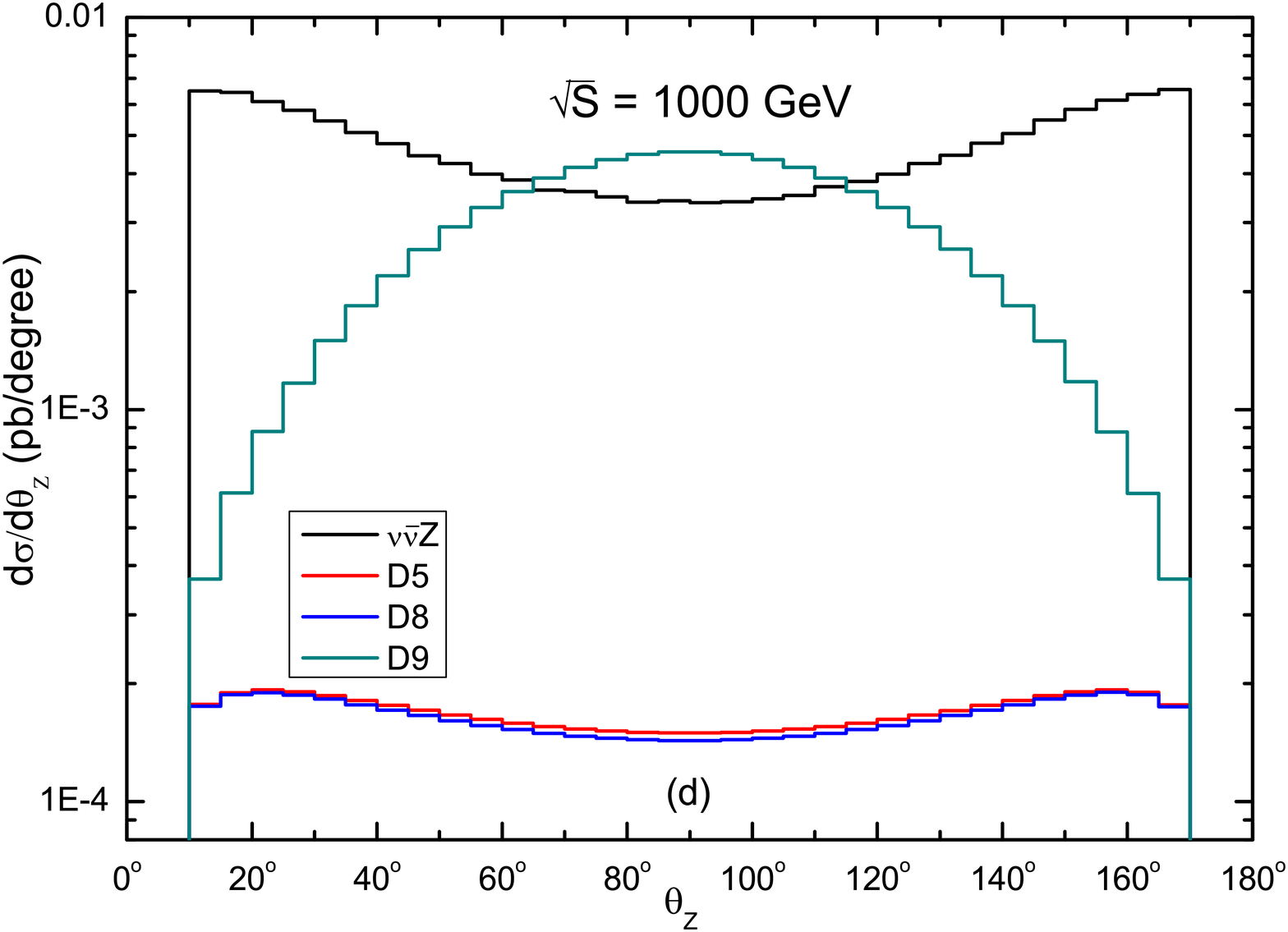}
\includegraphics[width=0.45\columnwidth]{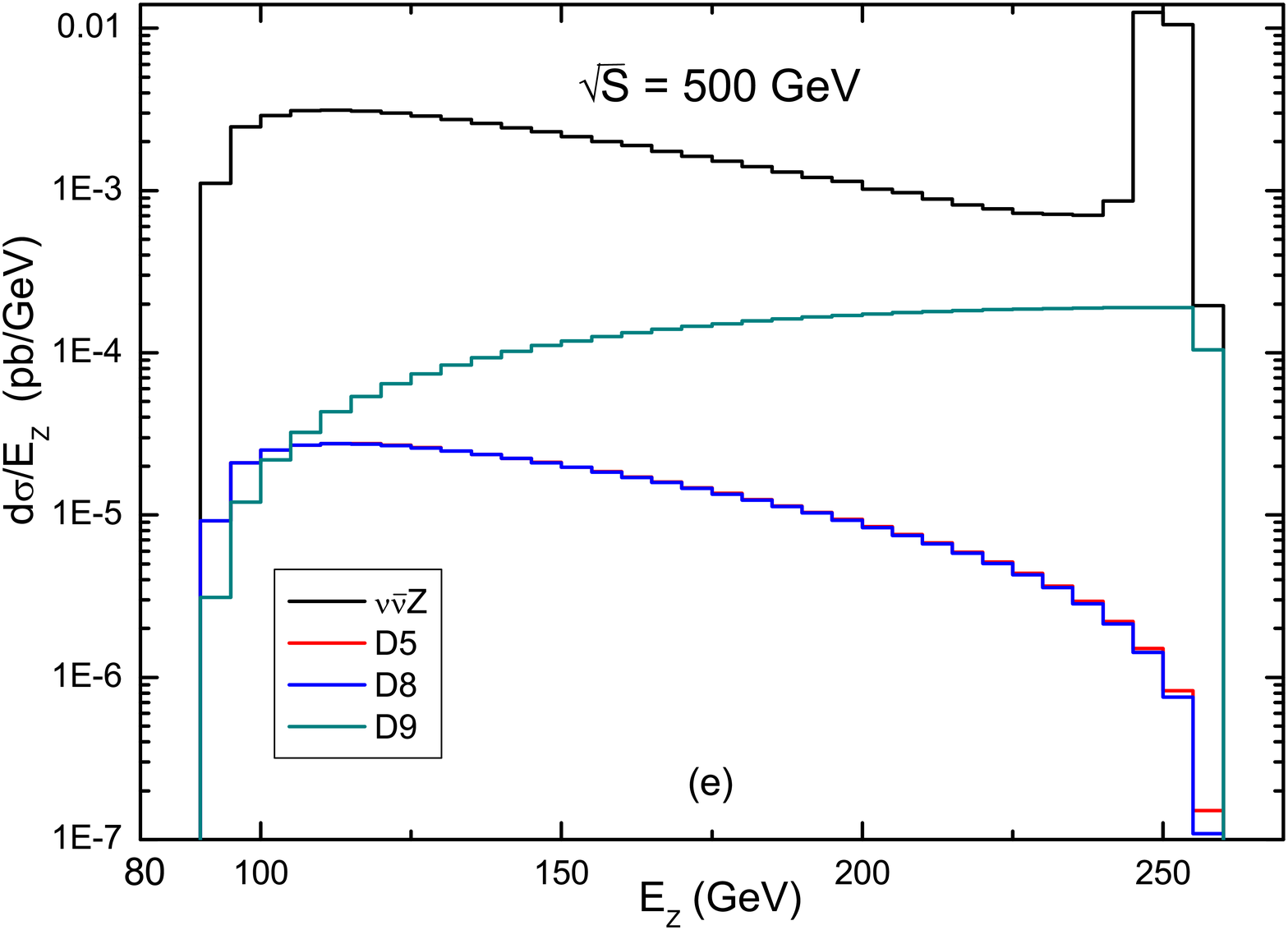}
\includegraphics[width=0.45\columnwidth]{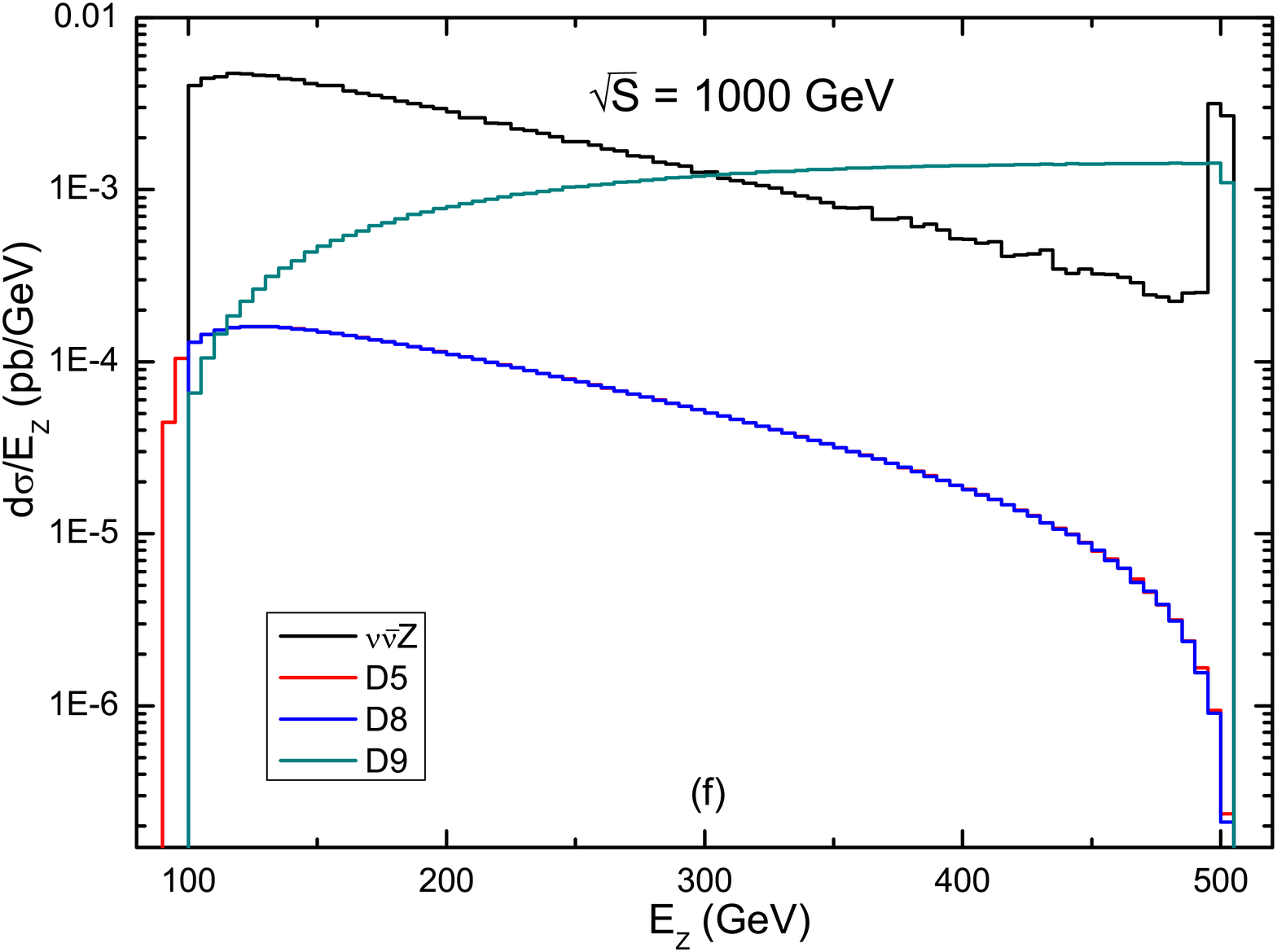}
\caption{ \label{fig5} Differential distributions of $p^Z_T$, $\theta_Z$ and $E_Z$
for the signal induced by the vector, axial-vector, tensor operators and the background at the
$\sqrt{s} = 500$ and $1000~ {\rm GeV}$ ILC with $m_\chi=10~{\rm GeV}$, $\Lambda=1~{\rm TeV}$ and
$10^{\circ} < \theta_z < 170^{\circ}$. }
\end{center}
\end{figure}

\par
The significance of signal over background $S$ is defined as
\begin{eqnarray}
S = \frac{N_S}{\sqrt{N_B}} = \frac{\sigma_S \sqrt{{\cal L}}}{\sqrt{\sigma_B}},
\end{eqnarray}
where $N_{S,B}$ and $\sigma_{S,B}$ are the event numbers and cross
sections for signal and background, and ${\cal L}$ denotes the
integrated luminosity. In Fig.\ref{fig4}, we depict the $3 \sigma$
detection region (defined as $S \geq 3$) on the $m_\chi-\Lambda$
plane by taking $\sqrt{s} = 500$ and $1000~ {\rm GeV}$, ${\cal L} =
100$ and $1000 fb^{-1}$ and $10^{\circ} < \theta_z < 170^{\circ}$,
respectively. As mentioned in Fig.\ref{fig2}, we know that $\Lambda$
has a larger space of adjustment by improving colliding energy.
Of course, we can also reach the same effectiveness with the improvement of integrated
luminosity. As shown Fig.\ref{fig4}, we can
see that $\Lambda$ will increase with the improvement of $\sqrt{s}$
or integrated luminosity. In Table \ref{tab1}, we list the cross
sections for the signal process $e^+e^- \rightarrow \chi\bar{\chi}
Z$ and the SM background at the $\sqrt{s} = 500,~ 1000~ {\rm GeV}$
ILC, and the corresponding significances with ${\cal L} = 100$ and
$1000~ fb^{-1}$, where the DM mass $m_{\chi}$ and the energy scale
$\Lambda$ are taken as $10~ {\rm GeV}$ and $1~ {\rm TeV}$,
respectively. Since the cross section for the signal is proportional
to $1/\Lambda^4$, we can transform the function of cross section
into a limit on the parameter $\Lambda$.

\par
In Fig.\ref{fig5} we present the transverse momentum ($p^Z_T$), angle ($\theta_Z$) and energy ($E_Z$)
distributions of the final visible $Z$ boson
for the signal induced by the vector,
axial-vector, tensor operators and background at the $\sqrt{s} =
500$ and $1000~{\rm GeV}$ ILC, respectively, with $m_\chi=10~{\rm
GeV}$, $\Lambda=1~{\rm TeV}$ and $10^{\circ} < \theta_z <
170^{\circ}$. The black, red, blue and green curves are for the SM
background and the signal induced by vector, axial-vector and tensor
operators, respectively. Our results show that the background is
very large compared to the signal. In order to efficiently
separate the $\chi \bar{\chi} Z$ signal from the $\nu \bar{\nu} Z$ background,
we need to adopt a proper event selection to
increase the signal-to-background ratio. For the
tensor(D9) interaction, the differential distribution of the signal
is very different from the background, and the differential cross
section is much larger than the contributions from D5 and D8.
By taking appropriate cuts, it is easily to separate the signal from
the background, and raise the limit on parameter $\Lambda_{D9}$.
However, the differential cross sections of the D5 and D8 are small
relative to that of D9, and the change of the signal distribution is
consistent with the background. It is difficult to distinguish the
signal from the background through the cuts of $p^Z_T$, $\theta_Z$,
$E_Z$.

\begin{table}[tbp]
\caption{Cross sections for the signal process $e^+e^- \rightarrow \chi\bar{\chi} Z$ and the SM background,
and the corresponding significances,
at the $\sqrt{s} = 500,~ 1000~ {\rm GeV}$ ILC with $m_\chi=10~{\rm GeV}$, $\Lambda=1~{\rm TeV}$,
$10^{\circ} < \theta_z < 170^{\circ}$ and two typical luminosity values of ${\cal L}_1 = 100~ fb^{-1}$ and
${\cal L}_2 = 1000~ fb^{-1}$. }
\label{tab1}%
\begin{ruledtabular}
\begin{tabular}{cccccc}
$\sqrt{s}~ ({\rm TeV})$ && $\sigma_S~ (fb)$ & $\sigma_B~ (fb)$ & $\sigma_S\sqrt{{\cal L}_1}/\sqrt{\sigma_B}$ & $\sigma_S \sqrt{{\cal L}_2}/\sqrt{\sigma_B}$ \\
\hline
                   &D5      &2.38             &                        &1.20             &3.80           \\
     0.5           &D8      &2.36            &392.31                   &1.19              &3.77          \\
                   &D9      &21.48             &                       &10.84              &34.29        \\
\hline
                   &D5      &27.13              &                      &9.96              &31.51         \\
      1.0          &D8      &26.35           &741.42                   &9.68              &30.60         \\
                   &D9      &423.18            &                       &155.42             &491.47       \\
\end{tabular}
\end{ruledtabular}
\end{table}

\vskip 5mm
\section{Summary}
\par
The origin of dark matter remains one of the most compelling mysteries in our understanding of the universe today.
High energy colliders are ideal facilities to search for DM. In this paper, we study the effects of the effective
operators of DM via dark matter pair production associated a $Z$ boson at the ILC.
The SM background $e^+e^- \rightarrow \nu_{\ell} \bar{\nu_{\ell}}Z$ $(\ell = e, \mu, \tau)$ is also considered for
comparison. We obtain the cross sections as functions of colliding energy $\sqrt{s}$ and the DM mass $m_\chi$ for
the signal induced by the vector, axial-vector, tensor operators and the SM background. We find that raising the
colliding energy can improve the probability for searching DM, and the contributions from the spin-independent operator
(D5) and the spin-dependent operator (D8) can be distinguished until $m_{\chi} > 100~{\rm GeV}$. If this signal is
not observed at the ILC, we set a lower limit on the new physics scale $\Lambda$ at the $3 \sigma$ level. Meanwhile,
we show the differential distributions of $p^Z_T$, $\theta_Z$ and $E_Z$. We find that it is easy to separate the
signal from the background by taking appropriate cuts for the tensor (D9) interaction, but difficult for the D5 and D8
interactions. We conclude that the ILC has the potential to detect the $e^+e^- \rightarrow \chi\bar{\chi} Z$ production.

\section{Acknowledgments}
This work was supported in part by the National Natural Science Foundation of China (No.11205003, No.11305001,
No.11275190, No.11375171, NO.11175001), the Key Research Foundation of Education
Ministry of Anhui Province of China (No.KJ2012A021), the Youth Foundation of Anhui Province(No.1308085QA07),
and financed by the 211 Project of Anhui University (No.02303319).


\end{document}